%% file: main.tex
\documentclass[10 pt, journal]{IEEEtran}  
\IEEEoverridecommandlockouts
\usepackage{cite}
\usepackage{amsmath,amssymb,amsfonts}
\usepackage{graphicx}
\usepackage{textcomp}
\usepackage{xcolor}
\usepackage{accents}
\usepackage[mathscr]{euscript}
\usepackage{mathrsfs}
\usepackage{mathtools}
\usepackage{textcomp, gensymb}
\usepackage{booktabs}
\usepackage{multirow}
\usepackage{makecell}
\usepackage{array}
\usepackage{algorithm, algpseudocode}
\usepackage{subcaption}
\usepackage[capitalize]{cleveref}
\usepackage{comment}

\newcommand\elb[1]{^{\left\{#1\right\}}}
\newcommand\nodelg[1]{{\mathcal{V}_{#1}\elb{i}}}

\begin{document}

\title{Optimality Loss Minimization in Distributed Control with Application to District Heating}

\author{Audrey~Blizard~and~Stephanie~Stockar%
\thanks{Audrey Blizard (blizard.1@osu.edu) and Stephanie Stockar (stockar.1@osu.edu) are with the  Department of Mechanical and Aerospace Engineering and the Center for Automotive Research, The Ohio State University, 930 Kinnear Road, Columbus, OH 43212 USA}}

\maketitle

\begin{abstract}
This paper presents a novel partitioning method designed to minimize control performance degradation resulting from partitioning a system for distributed control while maintaining the computational benefits of these methods. A game-theoretic performance metric, the modified Price of Anarchy, is introduced and is used in a generalizable partitioning metric to quantify optimality losses in a distributed controller. By finding the partition that minimizes the partitioning metric, the best-performing distributed control design is chosen. The presented partitioning metric is control-design agnostic, making it broadly applicable to many control design problems. In this paper, the developed metric is used to minimize the performance losses in the distributed control of a demand-flexible District Heating Network. The final distributed controller is provably feasible and stable. In simulation, this novel partitioning performed similarly to the centralized controller, increasing overall heat losses by only 1.9\%, as compared to a similarly-sized baseline partition, which resulted in a 22\% increase in losses.
\end{abstract}


\section{Introduction}
\IEEEPARstart{D}{istributed} control offers a practical solution for managing large-scale systems in a computationally efficient way \cite{christofidesDistributedModelPredictive2013}. By decomposing the centralized receding-horizon optimization problem into smaller subproblems, the issue of problem scale is eliminated. However, in the problem decomposition, key information can be lost, introducing an optimality gap between the distributed and centralized controllers. This gap is exacerbated in communication-based control where the actions of neighboring subsystems are treated as disturbances to the local subsystem \cite{camponogaraDistributedModelPredictive2002}. Choosing effective partitions is a promising approach to minimize the information losses and improve performance while still maintaining computation tractability \cite{sukhadeveEffectControlStructure2024,blizardOptimalityLossMinimization2025}.\par
District Heating Networks (DHNs) provide a compelling case for exploring effective partitioning in distributed control. DHNs consist of interconnected buildings that exchange heat through a network of underground pipes, making them large-scale, highly coupled systems.  Initial studies indicated that advanced control of DHNs can reduce network losses and peak heat demand \cite{vanoevelenTestingEvaluationSmart2023}. One of the main ways these performance benefits are seen is via the demand flexibility in the connected buildings \cite{vandermeulenControllingDistrictHeating2018}. In this control framework, the buildings are allowed to temporarily deviate from their nominal heat demand, allowing for more effective heat delivery in the network while still maintaining user comfort \cite{blizardUsingFlexibilityEnvelopes2024b}.\par
However, due to the large number of states, complex optimization objectives, and nonlinear dynamics, centralized controllers that consider the behavior of more than a few buildings are impractical in real-time \cite{vanoevelenTestingEvaluationSmart2023, delorenziPredictiveControlCombined2022}. Instead, distributed control has been explored as an effective method to decompose these large-scale control problems into computationally tractable subproblems \cite{zhaoOptimalControlHydraulic2023, caiAgentbasedDistributedDemand2020}. In this paper, we focus on communication-based distributed control, as this method has been shown to be highly effective in systems with strong state coupling \cite{scattoliniArchitecturesDistributedHierarchical2009}, such as DHNs. \par
The features that make centralized control of large-scale DHNs difficult also limit the effectiveness of existing system partitioning methods. The nonlinear and complex system dynamics make the coupling in the system model unclear and challenging to extract. Currently, most system partitioning is done based on ad-hoc decisions informed by the topology of the system; however, there are some existing partitioning methods based on system dynamics and coupling.\par
For one, interaction metrics give a measure of the strength of coupling between an input and an output, which are used to determine the groupings of the subsystems. The Relative Gain Array (RGA) measures the effect of a single input on a state compared when all inputs are active \cite{bristolNewMeasureInteraction1966,manousiouthakisSynthesisDecentralizedProcess1986}. It has also been adapted to account for the relative degree between an input and output and different time scales \cite{yinInputOutputPairing2017,tangRelativeTimeaveragedGain2018}. However, RGA-based techniques are ineffective for partitioning complex systems, as they fail to provide insight into systems where several outputs are influenced by a single input.
The Hankel Interaction Index Array (HIIA) measures the throughput of each state-input pair \cite{wittenmarkHankelnormBasedInteraction2002}. This technique was extended to include systems with long time delays, using the ratio of open- to closed-loop Hankel norms \cite{moaveniModifiedHankelInteraction2018}. While Hankel norm-based interaction measures consider the full range of dynamic performance of the system, they fail to consider the closed-loop properties of the resulting subsystems.\par
The other key partitioning method is graph-based partitioning, where a graph is created to represent the system. The graph's nodes are the states, inputs, and outputs of the system, and the edges represent the coupling between these nodes. In its most straightforward implementation, an unweighted, undirected equation graph is used, and edges represent the existence of coupling between two nodes \cite{jogwarCommunitybasedSynthesisDistributed2017}. In more recent implementations, the weight of each edge has been decided based on the coefficients of the linear state space model, \cite{jogwarDistributedControlArchitecture2019}, and with time-varying state-space models \cite{ebrahimiAdaptiveDistributedArchitecture2024}. This can also be supplemented with a binary term to represent whether states are dynamically or algebraically coupled \cite{tangNetworkDecompositionDistributed2018}. The created graphs are then partitioned using a variety of graph partitioning techniques \cite{jogwarDistributedControlArchitecture2019,ebrahimiAdaptiveDistributedArchitecture2024}. However, in DHNs, much of the system coupling and performance limitations are seen in the algebraic equality and inequality constraints in the problem, such as flow balancing and heat requirements. These factors are difficult to fully capture within this graph-based framework. \par
The existing system partitioning methods lack closed-loop performance considerations, leading to large performance degradation when designing a distributed controller. This paper proposes an alternative method that quantifies the closed-loop performance losses introduced when decomposing the system and then minimizes these performance losses. By interpreting the partitions as agents, each with local objectives that interact to achieve a global goal, game theory provides a robust framework for analyzing the closed-loop realizations of the system and their resulting performance. This novel perspective allows the proposed partitioning method to move beyond those based on traditional open-loop state coupling. The goal of introducing this novel performance-based partitioning is to design a scalable distributed controller that performs as well as the centralized version.\par
Previous work in multi-agent control using game theory has primarily focused on Shapley values for coalition formation, grouping agents into cooperative units working towards a common goal, rather than system partitioning \cite{murosPartitioningLargeScaleSystems2018}. Moreover, the method relies on the strong assumption that the system is controllable in a fully decentralized fashion, without any communication \cite{murosGameTheoreticalRandomized2018}, which is difficult in complex and highly coupled systems like DHNs. \par
The method developed in this paper determines the system partition that closes the optimality gap without increasing computation time, thereby providing a benchmark for the best possible non-cooperative controller. In addition to characterizing the performance losses from transitioning from a centralized to a distributed control structure, the presented method possesses several key properties. First, it is independent of the control approach and applicable to any optimal control strategy, such as a Linear Quadratic (LQ) Regulator or MPC that relies on communication to achieve a consensus between agents. Second, it accommodates different control objectives for each agent to exploit the advantages of a non-cooperative control structure. Additionally, it considers practical aspects affecting the controller’s implementability, such as partition size and convergence of the distributed controller. Finally, it ensures that the resulting partitions are stable and able to satisfy system constraints at any time step.\par
The remainder of the paper is organized as follows. First \cref{sec:olm} presents the general distributed control formulation, which is then used to develop the novel system partitioning method. Then, the DHN-specific distributed control problem is introduced in \cref{sec:dist_dhn}, along with guarantees of the feasibility and stability of the partitioned problem. \Cref{sec:case_study} presents the four-user case study considered in this paper, comparing the results to a baseline partitioning method. Finally, in \cref{sec:conc}, conclusions and future work are presented. For clarity, the notation used in this paper is contained in \cref{sec:notation}.

\section{Optimality Loss Minimization}\label{sec:olm}
\subsection{Distributed Control Formulation}\label{sec:dist_gen}
Consider a general nonlinear control problem with states $x\in\mathbb{R}^{n_x}$ and control input $u\in\mathbb{R}^{n_u}$ written as
\begin{subequations}\label{eq:cen}
    \begin{equation}
        c_{cen} = \min_{u} \sum_{k=t_0}^{t_0+\Delta t} f(x,u)\ s.t.
    \end{equation}
    \begin{equation}
        g(x,u)\leq 0
    \end{equation}
    \begin{equation}\label{eq:dynamics}
        x(k+1) = h(x,u)
    \end{equation}
\end{subequations}
where $f(x,u)$ is the optimization objective, $g(x,u)$ gives the constraints on the problem, and \cref{eq:dynamics} governs the discrete state evolution of the system from an initial condition. Additionally, $t_0$ is the initial time and $\Delta t$ is the optimization horizon. Solving this problem gives the centralized optimal control action for the system over the control horizon.\par
This optimization problem is represented by a graph $\mathcal{G} = (\mathcal{V},\mathcal{E})$, where $\mathcal{V}=\left\{x,u\right\}$ is the set of state and control variables. The edges represent any direct connections between these variables in $g$ and $h$. From this perspective, the conversion of a centralized optimization problem to a communication-based distributed one is considered as a graph partitioning problem, where the nodes in $\mathcal{V}$ are separated into disjoint groups as
\begin{equation}
\begin{split}
    \operatorname{part}\left(\mathcal{V}\right) &= \left\{\mathcal{V}\elb{1},\dots,\mathcal{V}\elb{n_p}\right\}\\
    \text{ s.t. } \mathcal{V} &= \bigcup_{i = 1}^{n_p} \mathcal{V}\elb{i},\quad
    \mathcal{V}\elb{i}\cap \mathcal{V}\elb{j}=\emptyset\ \forall i\neq j
    \end{split}
\end{equation}
where $\mathcal{V}\elb{i}$ is a single partition, and $n_p$ is the total number of partitions.\par In this control framework, communication is used to compensate for any cut edges.
This communication is represented by the set of edges connecting nodes from different partitions,
\begin{equation}
    \mathcal{E}_C = \left\{\ e\left(v_j,v_i\right) \mid v_i\in\mathcal{V}\elb{i},\ v_j\in\mathcal{V}\elb{j},\ i\neq j\ \right\}
\end{equation}
where $e\left(v_j,v_i\right)$ is an edge directed from $v_j$ to $v_i$. From $\mathcal{E}_C$, the set of neighbors for a partition $\mathcal{V}\elb{i}$ is defined as
\begin{equation}
    \mathcal{N}\elb{i} = \left\{ v_j \mid e\left(v_j,v_i\right)\in \mathcal{E}_C \right\}
\end{equation}
The neighbors are used to find the local constraint function
\begin{equation}
\begin{gathered}
g\elb{i}(x\elb{i},u\elb{i},x\elb{j},u\elb{j})\\
x\elb{i},u\elb{i}\in \mathcal{V}\elb{i} \text{ and } x\elb{j},u\elb{j}\in \mathcal{N}\elb{i}
\end{gathered}
\end{equation}
and state equation
\begin{equation}
    \begin{gathered}
        h\elb{i}(x\elb{i},u\elb{i},x\elb{j},u\elb{j})\\
        x\elb{i},u\elb{i}\in \mathcal{V}\elb{i} \text{ and } x\elb{j},u\elb{j}\in \mathcal{N}\elb{i}
    \end{gathered}
\end{equation}
which includes the values of the disturbances caused by neighboring subsystems. 
In non-cooperative, communication-based distributed control, a subsystem's objective function is
\begin{equation}
    f\elb{i}(x\elb{i},u\elb{i})
\end{equation}
only considering local variables, which is preferable in large-scale systems, as a global cost function would require too much information transmission. However, without a global cost function, the system may not converge to the cooperative globally optimal solution. Instead, the solution to a non-cooperative, communication-based distributed control problem is a Nash equilibrium (NE) where no subsystem can unilaterally improve its performance \cite{duDistributedModelPredictive2001}.
This NE is found according to
\begin{subequations}\label{eq:dist}
    \begin{equation}
        c\elb{i} = \min_{u\elb{i}} f\elb{i}(x\elb{i},u\elb{i})\ s.t.
    \end{equation}
    \begin{equation}
        g\elb{i}(x\elb{i},u\elb{i},x\elb{j}(p),u\elb{j}(p))\leq 0
    \end{equation}
    \begin{equation}
        \dot{x}\elb{i} = h(x\elb{i},u\elb{i},x\elb{j}(p),u\elb{j}(p))
    \end{equation}
\end{subequations}
where $x\elb{j}(p)$ and $u\elb{j}(p)$ are the predicted trajectories of the neighboring subsystems, updated after each solution iteration until a consensus is achieved. \par
This NE is often not socially optimal, as the individual agents behave selfishly, without considering overall system performance \cite{duDistributedModelPredictive2001}, and the cost associated with an NE is always greater than or equal to the centralized optimal cost \cite{blizardOptimalityLossMinimization2025}. \par
\subsection{Optimality Loss Minimization}
The system partitioning problem is posed as an optimization problem, minimizing the optimality losses incurred by converting the problem to a distributed one while maintaining computational efficiency through sufficiently small subsystems. The performance degradation caused by the selfish behavior of the agents in a non-cooperative problem is quantified by the Price of Anarchy (PoA)
given by 
\begin{equation}
    c_{PoA} = \frac{\min_{NE}\sum_{i=1}^{n_p}{c_{NE}\elb{i}}}{c_{cen}}
\end{equation}
where $c_{cen}$ is the cost associated with the centralized solution, and $\min_{NE}\sum_{i=1}^{n_p}{c_{NE}\elb{i}}$ is the total cost of all subsystems associated with the worst-performing NE \cite{koutsoupiasWorstcaseEquilibria2009}. This game-theoretic concept can be calculated for any system of agents that converges to an NE, making it a generalizable metric for the distributed control partitioning problem. \par
However, the worst-case NE is easily escaped, \cite{seatonIntrinsicFragilityPrice2023}, making it an unfairly pessimistic metric for quantifying a distributed controller's performance. Instead, a modified Price of Anarchy (mPoA) metric, defined as
\begin{equation}
    c_{mPoA} = \frac{\sum_{i=1}^{n_p}{c_{NE_s}\elb{i}}}{c_{cen}}
\end{equation}
is used, where ${NE_s}$ is the found stable NE solution assuming a standard initial guess for the predicted trajectories.
This mPoA is used to develop the optimal partitioning problem given by
\begin{subequations}\label{eq:olm}
    \begin{equation}
        \min_{\operatorname{part}\left(\mathcal{V}\right)} olm\ s.t.
    \end{equation}
    \begin{equation}
    olm = w_1c_{mPoA}+w_2c_{iter}+w_3c_{sz}
    \end{equation}
\end{subequations}
The optimality loss metric (OLM) includes three terms. The first term is the mPoA, where $c_{cen}$ is found using \cref{eq:cen} and $c\elb{i}$ according to \cref{eq:dist}. Additionally, the OLM includes the size of the largest partition $c_{sz} = \max_i \left(\operatorname{card}\mathcal{V}\elb{i}\right)$, to balance the optimality losses with computation time and encourage further partitioning. Note that only the size of the largest partition is considered, as the local optimization will be solved in parallel, so the complexity of the largest subproblem dictates computation time. As the solution depends on iterative communication to reach a consensus, the number of iterations needed will directly impact solving time. Therefore, the OLM also includes the number of iterations to convergence ($c_{iter}$), to choose the fastest converging result, given similar size and mPoA values. Finally, $w_i,\ i=1 \dots 3$ are weighting terms, with decreasing priority to reflect the importance of each factor in the optimization process. \par

\section{Distributed Control of DHNs}\label{sec:dist_dhn}
\begin{figure}
    \centering
    \includegraphics[width=\linewidth]{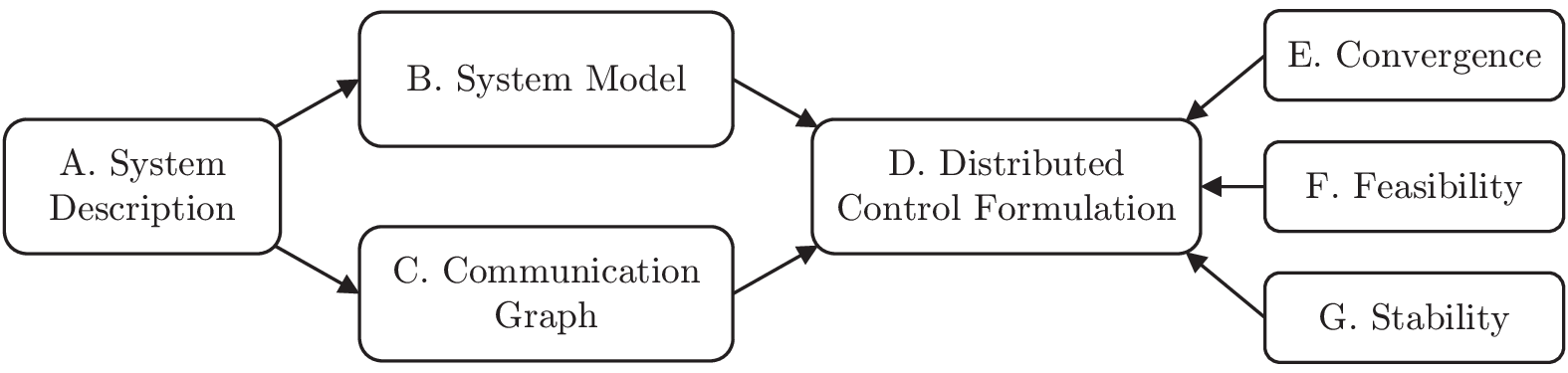}
    \caption{Visual roadmap of \cref{sec:dist_dhn} to aid comprehension of how each subsection contributes to the distributed control design.}
    \label{fig:roadmap}
\end{figure}
One of the benefits of the presented partitioning algorithm is that it works for any communication-based distributed control method. In this paper, the nonlinear communication-based distributed control method presented and validated for effectiveness  \cite{blizardCommunicationBasedDistributedControl2025} will be used to optimize the behavior of a demand-flexible DHN. The formulation and communication structure of the distributed controller will be described in this section. An overview of the flow of this section is presented in \cref{fig:roadmap}.
\subsection{System Description}
\begin{figure}\label{fig:graphs}
    \centering
    \begin{subfigure}[t]{.5\linewidth}
        \centering
        \includegraphics[width=1\linewidth,trim={35 20 25 15},clip]{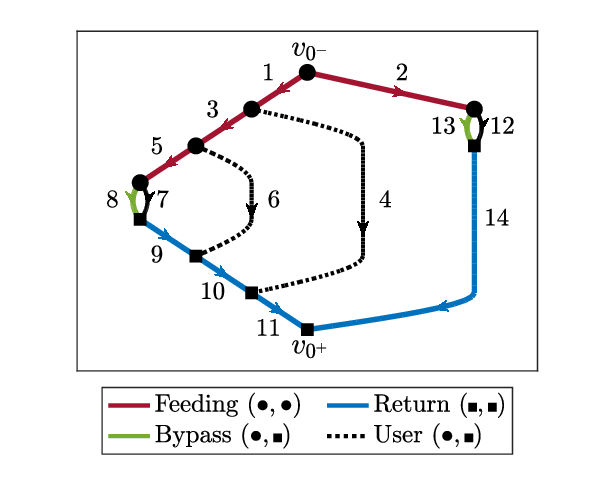}
        \caption{Flow graph.}
        \label{fig:flow_graph}
    \end{subfigure}%
    \begin{subfigure}[t]{.5\linewidth}
        \centering
        \includegraphics[width=1\linewidth,trim={35 20 25 15},clip]{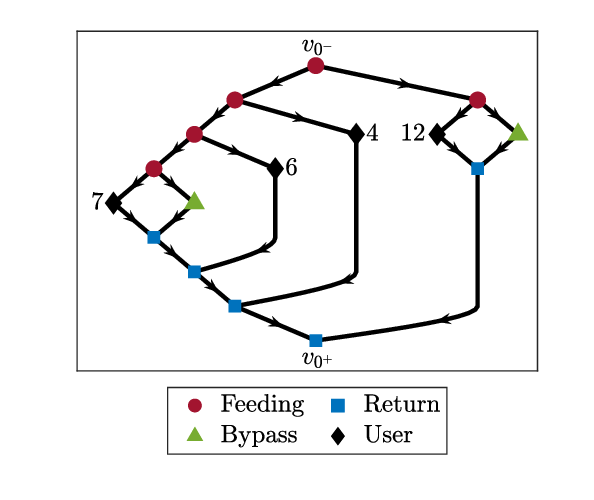}
        \caption{Line graph.}
        \label{fig:line_graph}
    \end{subfigure}
    \caption{Modeling graphs for the four-user case study.}
\end{figure}

The flow of the DHN is represented by an unweighted rooted directed graph $\mathcal{G}_f = (\mathcal{V}_f,\mathcal{E}_f)$ where the edges $\mathcal{E}_f$ represent the network components, and the nodes are where the flows are split and recombined. Additionally, the plant is modeled as two nodes $\{v_{0^-}, v_{0^+}\}$, the root and terminal nodes of the graph, respectively. The connections between the nodes and edges in $\mathcal{G}_f$ are represented by an incidence matrix $\Lambda$, which encodes the in and out nodes of each edge, which will be used in the flow calculations of the network. The flow graph for the four-user case study is presented in \cref{fig:flow_graph}.\par
Additionally, this representation also correlates to a line graph, where all of the edges of the original graph are represented by nodes, and the edges represent connections between them. An example is shown in \cref{fig:line_graph}. This graph is given by $\mathcal{G} = (\mathcal{V},\mathcal{E})$ where the nodes $\mathcal{V}$ represent the network components and is composed of four disjoint subsets, $\mathcal{V}=\{F, R,{By},{U}\}$ for the feeding lines, return lines, bypass lines, and user elements, respectively. $\mathcal{E}$ is the set of edges indicating connections between these elements. Additionally, the plant is included as two additional nodes $\{v_{0^-}, v_{0^+}\}$. These node sets, with the adjacency matrix of $\mathcal{G}$, $\Gamma$, identify relationships between the pipes.\par
The goal of the optimal control problem is to use the users connected to the network as flexible consumers of heat to minimize heat losses in the network. The control variables are the mass flow rate supplied to the plant $\dot{m}_0$, and the valve positions of the users $\theta$, which control the flow to the users and therefore the heat delivered. The external disturbances are the ambient temperature $T_{amb}$, the nominal heat demand of the users $\dot{Q}_{amb}\elb{U}$, and the supply temperature $T_{0}$. The nominal heat demand of the users is assumed to be a known constant, as only small temperature deviations will be allowed. Any nodes connected to $v_{0^-}$ will receive water with the predefined supply temperature. \par
\subsection{System Model}\label{sec:model}
The thermal and flow behaviors of the network are modeled using first-principles methods as follows.
\subsubsection{Thermal Model}
To model the behavior of an individual pipe segment, a well-mixed approach is used, where the bulk temperature in a pipe, $T_P\elb{i}$, is found according to 
\begin{equation}\label{eq:Tpipe}
    \rho c_pV_P\elb{i} \frac{d}{dt}T_P\elb{i} = \dot{Q}_{in}\elb{i}+\dot{Q}_{amb}\elb{i}
\end{equation}
where $\rho$ and $c_p$ are the operating fluid's density and specific heat capacity, respectively, and $V_P$ is the pipe's volume. The heat transferred into the pipe by the upstream flow $\dot{Q}_{in}\elb{i}$ is given by
\begin{equation}\label{eq:Qin}
    \dot{Q}_{in}\elb{i} = \dot{m}_P\elb{i}c_p\left(T_P\elb{i^-}-T_P\elb{i}\right)
\end{equation}
where $\smash{\dot{m}_P\elb{i}}$ is the current pipe's mass flow rate, and $\smash{T_P\elb{i^-}}$ is the temperature of the flow into the pipe.
The heat lost to the environment $\dot{Q}_{amb}\elb{i}$ is given by
\begin{equation}\label{eq:Qamb}
    \dot{Q}_{amb}\elb{i} = \left(hA_s\right)\elb{i}\left(T_{amb}-T_P\elb{i}\right)
\end{equation}
where $\smash{\left(hA_s\right)\elb{i}}$ is the heat transfer coefficient of the pipe.
Using \cref{eq:Tpipe} and $\Gamma$, the single pipe equations are combined to form a bilinear network model of the form
\begin{equation}
\label{eq:AugSS}
    \frac{d}{dt}T_P\elb{N} = A\left(\dot{m}_P\right)T_P\elb{N} +E\left(\dot{m}_P\right) \begin{bmatrix}T_0&T_{setR}&T_{amb}\end{bmatrix}
\end{equation}
where $N=F\cup R\cup By$, and $T_P\elb{N}$ is the vector of temperatures for these pipes. The $A$ and $E$ matrices are generated from the coefficients in \cref{eq:Tpipe} based on the inlet temperatures to each pipe, encoded in $\Gamma$ \cite{blizardGraphBasedTechniqueAutomated2024}.
Additionally, as the heat transfer out of $U$ is much faster than the rest of the network's dynamics, the temperatures in these edges are considered constant, $T_{setR}$, met by modulating the fluid's recirculation speed in the heat exchanger.
\subsubsection{Fluids Model}
An algebraic flow model is used to calculate the mass flow rates in the pipes. 
The pressure loss, $\Delta P_P\elb{i}$, in an individual pipe segment is modeled by
\begin{equation}
\label{eq:dP}
    \Delta P_{P}\elb{i} = \zeta\elb{i}\left(\dot{m}_{P}\elb{i}\right)^2
\end{equation}
where $\zeta$ is the total pressure loss coefficient, determined based on the pipe's characteristics, including length, diameter, and material. In the user edges, $\zeta\elb{U}$ is a function of the valve positions $\theta_{min}<\theta<1$ given by
\begin{equation}\label{eq:valve}
    \zeta\elb{U} = \mu\left(\theta^{-1}-1\right)^2
\end{equation}
where $\theta_{min}$ is set by the minimum pressure loss and $\mu$ is a scaling coefficient.\par
The conservation of mass in the network is enforced by
\begin{equation}
    \Lambda \dot{m}_{P} = \dot{m}_{S}
\end{equation}
where $\dot{m}_P$ is the vector of mass flow rates in the pipes and $\dot{m}_S$ is the vector of the mass flow rates in the network splits, given by
\begin{equation}
    \dot{m}_S\elb{i} =\begin{cases}
    \dot{m}_0 & \text{if } v_i = v_{0^+}\\
    -\dot{m}_0 & \text{if } v_i = v_{0^-}\\
    0 &\text{otherwise}
    \end{cases}
\end{equation}
The pressure balance in the network is found using 
\begin{equation}
    \Delta P_{P} = \Lambda^\top  P_{S}, \quad P_{S}\elb{v_{0^-}}=0
\end{equation}
where $\Delta P_{P}$ is the vector of pressure losses from \cref{eq:dP} and $P_{S}$ is a vector of the pressure at every split, where $v_{0^-}$ serves as the reference. 
\subsubsection{Flexible Building Model}
The buildings are modeled using a flexibility envelope-based approach \cite{reyndersGenericCharacterizationMethod2017}, where the states of energy $SOE\elb{U}$ of the buildings are given by
\begin{equation}\label{eq:opt_soe}
    \frac{d}{dt}SOE\elb{U} = \dot{Q}_{in}\elb{U}-\dot{Q}_{amb}\elb{U}
\end{equation}
where the heat provided to the buildings $\dot{Q}_{in}\elb{U}$ are
\begin{equation}\label{eq:opt_qp}
    \dot{Q}_{in}\elb{U} = \dot{m}_P\elb{U}c_p\left(T_P\elb{U^-}-T_{setR}\right)
\end{equation}
where $U^-$ are the users' inlet edges. Additionally, associated with the buildings are heat storage capacity coefficients $C\elb{U}$ and acceptable temperature deviations $\Delta T_b$.

\subsection{Communication Graph}
As each pipe has 3 associated variables that interact with other pipes: temperature, mass flow rate, and pressure, there will be three sets of communication edges $\mathcal{E} = \left\{\mathcal{E}_T,\mathcal{E}_{\dot{m}},\mathcal{E}_{\Delta P}\right\}$ respectively. As the communication relies on the pipe states, the line graph $\mathcal{G}$ will be used to develop the communication graph.
Passing the network temperatures, node pressures, and node flow rates ensures consistency between the local subsystem solutions and the global network behavior. Each variable type is given a different passing direction to ensure each subsystem has at least one degree of freedom in its local optimization problem. Let $v_1$ be the pipe directly upstream from $v_2$. Temperature is transmitted consistent with the flow direction, so that 
\begin{equation}
    e \left(v_1,v_2\right) \in \mathcal{E}_T\\
\end{equation}
Additionally, the prescribed split pressures $P_S$ are passed between subsystems to ensure consistency in the pressure losses. First, each node pressure is associated with its downstream edge in $\mathcal{G}_f$. The communication edges are then assigned as
\begin{gather}
    v_1\in F,\ v_2\in F\cup U\cup By \Rightarrow e \left(v_1, v_2\right) \in \mathcal{E}_{\Delta P}\\
     v_1\in R\cup U\cup By,\ v_2\in R \Rightarrow e \left(v_2, v_1\right) \in \mathcal{E}_{\Delta P}
\end{gather}
The communication edges in $\mathcal{E}_{\dot{m}}$ are directed inversely of those in $\mathcal{E}_{\Delta P}$, so that
\begin{gather}
    v_1\in F,\ v_2\in F\cup U\cup By \Rightarrow e \left(v_2, v_2\right) \in \mathcal{E}_{\dot{m}}\\
     v_1\in R\cup U\cup By,\ v_2\in R \Rightarrow e \left(v_1, v_2\right) \in \mathcal{E}_{\dot{m}}
\end{gather}
Note that no edges exist between $v_1\in U, By$ and $v_2\in U, By$. Additionally, 
for $v_1\in F$ connected to the plant root node $v_{0^-}$ and $v_2\in R$ connected to the plant terminal nodes $v_{0^+}$ the connections are
\begin{gather}
    e\left(v_{0^-}, v_1\right) \in \mathcal{E}_T, \mathcal{E}_{\Delta P}, \quad e\left(v_1, v_{0^-}\right) \in \mathcal{E}_{\dot{m}}\\
    e \left(v_2, v_{0^+}\right) \in \mathcal{E}_T, \mathcal{E}_{\dot{m}}, \quad e \left(v_{0^+}, v_2\right) \in \mathcal{E}_{\Delta P}
\end{gather}\par
\begin{figure}
    \centering
    \includegraphics[width=.7\linewidth,trim={35 20 25 15},clip]{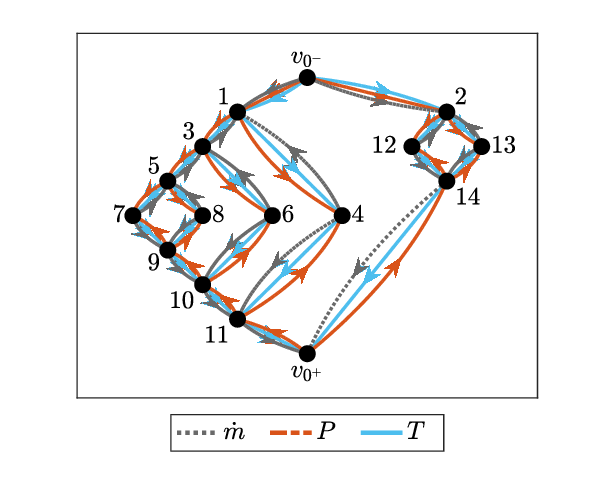}
    \caption{Communication graph.}
    \label{fig:comm_graph}
\end{figure}

The communication graph for the system in \cref{fig:graphs} is presented in \cref{fig:comm_graph}.
The created communication graph is partitioned as described in \cref{sec:dist_gen} to get the variables that must be communicated by the distributed controller. Note that $v_{0^+}$ is always assigned to a separate partition, as the only information passed out of it is reference pressure. Additionally, $v_{0^-}$ must always be included in a partition with at least one connected node so that the plant pressure is deterministic. \par
After the partitions are created, the neighboring set is divided into two, one upstream of the partition $\mathcal{N}\elb{i}_u$ and one downstream of the partition $\mathcal{N}\elb{i}_d$. 
In a slight abuse of notation, these sets are also used to denote the nodes that are connected to the edges connecting these neighbors.
\subsection{Distributed Control Formulation}
After the system is partitioned, the local, discrete-time optimization problem to be solved by each subnetwork is given by
\begin{subequations}
    \begin{multline}\label{eq:i_cost} 
       c\elb{i} = \min_{\mathclap{\widetilde{\dot{m}}_0,\theta\elb{U\elb{i}}}}\quad    \sum_{t_0}^{t_0+\Delta t} f \left( \dot{m}_{P}\elb{\mathcal{V}\elb{i}}, T_P\elb{N\elb{i}}, SOE\elb{U\elb{i}} \right)\\
    \text{subject to:}
    \end{multline}
    \begin{equation}\label{eq:i_temp}
           T_P\elb{N\elb{i}}(k+1)  = A\elb{i}T_P\elb{N\elb{i}} +E\elb{i} \begin{bmatrix}\widetilde{T}_{0}\\T_{setR}\\T_{amb}\end{bmatrix}
    \end{equation}
    \begin{equation}\label{eq:i_soe}
        SOE\elb{U\elb{i}}(k+1) = \dot{Q}_p\elb{U\elb{i}}-\dot{Q}_{out}\elb{U\elb{i}}
    \end{equation}
    \begin{equation}\label{eq:i_qp}
    \dot{Q}_p\elb{U\elb{i}} = \dot{m}_P\elb{U\elb{i}}c_p\left(T_P\elb{{U\elb{i}}^-}-T_{setR}\right)
    \end{equation}
    \begin{equation}\label{eq:i_dP}
        \Delta P_P\elb{\mathcal{V}\elb{i}} = \zeta\elb{\mathcal{V}\elb{i}} \left(\dot{m}_P\elb{\mathcal{V}\elb{i}}\right)^2
    \end{equation}
    \begin{equation}\label{eq:i_valve}
        \zeta\elb{U\elb{i}} = \mu\left({\theta\elb{U\elb{i}}}^{-1}-1\right)^2
    \end{equation}
    \begin{equation}\label{eq:i_flow}
        \Lambda_i \dot{m}_P\elb{\mathcal{V}\elb{i}} = \widetilde{\dot{m}}_S
    \end{equation}
    \begin{equation}\label{eq:i_pressure}
    \Delta P_P\elb{\mathcal{V}\elb{i}} = \Lambda_i^\top P_S\elb{\mathcal{E}\elb{i}}
    \end{equation}
    \begin{equation}\label{eq:i_flex}
         -C\elb{U\elb{i}}\Delta {T_b}\leq SOE\elb{U\elb{i}}\leq C\elb{U\elb{i}}\Delta {T_b}
    \end{equation}
    \begin{equation}\label{eq:i_theta}
        \theta_{min}<\theta\elb{U\elb{i}}<1
    \end{equation}
\end{subequations}
where \cref{eq:i_dP,eq:i_flex,eq:i_soe,eq:i_qp,eq:i_valve,eq:i_theta} are taken directly from \cref{sec:model}, considering only local variables.\par
In \cref{eq:i_temp}, $\widetilde{T}_0$ is the local inlet temperatures given by 
\begin{equation}
    \widetilde{T}_0 = \begin{bmatrix}
        T_0 & \text{if } v_{0^-}\in\mathcal{V}\elb{i}\\
        T_P\elb{\mathcal{N}\elb{i}_{T}}(p) & \text{if } \mathcal{N}\elb{i}_{T} \neq \emptyset
    \end{bmatrix}
\end{equation}
and $A\elb{i}$ and $B\elb{i}$ are subsets of the originals, and are functions of the local and neighboring mass flow rates. The vector of node mass flow rates in \cref{eq:i_flow}, $\widetilde{\dot{m}}_S$, is now given by 
\begin{equation}
    \widetilde{\dot{m}}_S\elb{v} =\begin{cases}
    \dot{m}_{i^+}\elb{v} & \text{if } v = v_{0^+}\\
    \dot{m}_{i^-}\elb{v} & \text{if } v = v_{0^-}\\
    \dot{m}_{i^-}\elb{v} & \text{if } v\in \mathcal{N}\elb{j}_{\dot{m}_{u}}\\
     \dot{m}_{i^+}\elb{v} & \text{if } v\in \mathcal{N}\elb{j}_{\dot{m}_{d}}\\
     \dot{m}_{j^+}\elb{v}(p) & \text{if } v\in \mathcal{N}\elb{i}_{\dot{m}_{u}}\\
     \dot{m}_{j^-}\elb{v}(p)& \text{if } v\in \mathcal{N}\elb{i}_{\dot{m}_{d}}\\
    0 &\text{otherwise}
    \end{cases}
\end{equation}
where $\dot{m}_{i^+}\geq0$ is a vector of controllable inlet flows, $\dot{m}_{i^-}\leq0$ is a vector of controllable outlet flows, and $\dot{m}_{j^-},\dot{m}_{j^+}$ are the flows controlled by the associated neighboring system.
Additionally, the controllable mass flow rate variables are combined into the control variable as
\begin{equation}
    \widetilde{\dot{m}}_0 = \begin{bmatrix}
        \dot{m}_{i^+}^\top&
        \dot{m}_{i^-}^\top
    \end{bmatrix}^\top
\end{equation}
This information is also used to recover the centralized plant mass flow $\dot{m}_0$ as
\begin{equation}
    \dot{m}_0 = \sum_{\mathclap{i=1\dots n_p\text{ s.t.}
    v_{0^+}\in\mathcal{V}\elb{i}}} \dot{m}_{i^+}\elb{v_{0^+}}(p)
\end{equation}
In subnetworks where any node pressures are dictated by a different subsystem, this pressure constraint is accounted for in \cref{eq:i_pressure} by setting additional split pressures according to
\begin{equation}\label{eq:constrP}
    P_S\elb{v} =\begin{cases}
        
    P\elb{v}_S(p) & \text{if } v \in\mathcal{N}_{\Delta P}\elb{i}\\
    0 & \text{if } v = v_{0^-}
    \end{cases} 
\end{equation}
where again the split pressure is associated with its downstream edge in the flow graph.
\subsection{Convergence}
Each subsystem solves the local optimization problem, assuming constant predicted values of the neighboring subsystems. After each solution, local variables 
\begin{equation}
    x_{\mathcal{N}} = \begin{bmatrix}
        {T_P\elb{\mathcal{N}\elb{j}_{T}}}^\top &
        {P_S\elb{\mathcal{N}_{\Delta P}\elb{j}}}^\top &
        {\widetilde{\dot{m}}_S{\elb{\mathcal{N}_{\dot{m}}\elb{j}}}}^\top
    \end{bmatrix}^\top
\end{equation}
are passed to neighboring subsystems. The predicted trajectories are then updated according to 
\begin{equation}
    x_{\mathcal{N}}\left(p+1\right) = \omega x_{\mathcal{N}}\left(p\right) + (1-\omega) x_{\mathcal{N}}
\end{equation}
where $\omega$ is the step size, and $p$ is the predicted solution. This process is iterated until convergence.\par
Two convergence criteria exist. First, the change in the passed variables and the subsystem cost meet
\begin{equation}
    \begin{bmatrix} x_{\mathcal{N}}\left(p\right)-x_{\mathcal{N}}\\ 
    c\elb{i}\left(p\right)-c\elb{i}
    \end{bmatrix} \leq \epsilon
\end{equation}
where $\epsilon$ is a vector of threshold values. Additionally, any subsystems in $\mathcal{N}\elb{i}$ must have also converged. Therefore, subsystems with limited dependent neighbors converge first, allowing dependent subsystems to follow. Solutions that depend on each other must converge at the same time.

\subsection{Feasibility}
If the optimization problem is solved for a single partition, a feasible solution is always available, assuming
\begin{enumerate}
    \item $C\elb{U}\Delta {T_b}$ is constant or increasing.
    \item An algebraic model of the heat exchanger and unconstrained plant flow, which  ensures unlimited $\dot{Q}_p\elb{U}$ 
    \item Solar irradiation and heat transfer into buildings are neglected.
\end{enumerate}
From these assumptions, it can be shown that if the building's heat capacity starts within the flexibility envelope, it will be able to remain within the flexibility envelope. However, if the optimization problem has more than one partition, a single subsystem will dictate $P_S\elb{v_{0^+}}$, removing assumption 2. This limit can mean that the required heat is not available to prevent a building from exceeding the lower limit of its flexibility envelope. To prevent infeasibility, a feasibility recovery algorithm is implemented, which allows other subsystems to establish a lower limit on the plant pressure, effectively restoring the unlimited flow assumption. Any subsystems where the building's heat demand can not be met temporarily solves an optimization with \cref{eq:i_cost} as
\begin{equation}
     \min_{\widetilde{\dot{m}},\theta\elb{U_i}} P_{slack}
\end{equation}
where $P_{slack}$ is the increase from the dictated plant pressure, and the constraint established in \cref{eq:i_pressure} is relaxed to 
\begin{equation}
    P_S\elb{v_{0^+}}\left(p\right) \rightarrow  P_{min}=P_S\elb{v_{0^+}}\left(p\right)+P_{slack}
\end{equation}
By minimizing the added slack, the minimum plant pressure needed to ensure feasibility, $P_{min}$,  is found. This lower pressure limit is then transmitted to the controlling subsystem, where the constraint
\begin{equation}
    P_S\elb{v_{0^+}}\geq P_{min}
\end{equation}
is added to ensure this feasibility is not violated. The distributed controller then functions as designed with the new constraint activated. Implementing this feasibility restoration guarantees the overall controller's feasibility regardless of the chosen partitioning.
\subsection{Stability}
Another key concern in selecting system partitions is stability. Here, the stability of the partitioned DHN is demonstrated regardless of the chosen partition to simplify the search for the OLM-minimizing partition. The stability of the partitioned DHN is formally proven in \cref{apx:stability}, and it is constructed using the passive nature of DHNs. First, the passive nature of a single element is demonstrated. Then, the passivity of the interconnected elements is shown. Finally, the zero-state detectability of the DHN is used to ensure stability results from the passive nature. 

\section{Simulation Study}\label{sec:case_study}
\subsection{Case Description}
The OLM minimizing partitioning will be looked at for the four-user network shown in \cref{fig:graphs}. Including pipes, the plant, and the users, this system has a total of 15 elements to be partitioned. The network was constructed using representative network parameters summarized in \cref{tbl:network} \cite{blizardDynamicallySimilarLabscale2023}. The network supply temperature, $T_0$ was assumed to be constant. The ambient temperature in typical meteorological year 3 (TMY3) was provided by NREL in the 2021 data set\cite{horowitz2019resstock}.\par
\begin{figure}
    \centering
    \includegraphics[width=.8\linewidth,trim={15 0 30 10},clip]{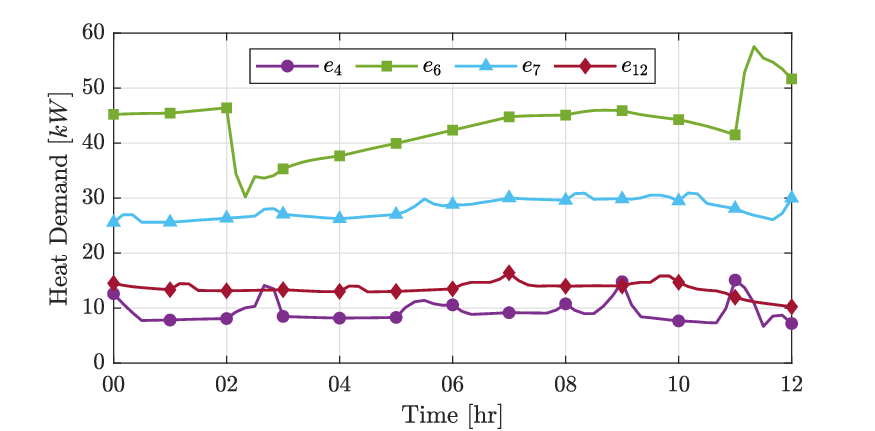}
    \caption{Nominal heat demand of the four users.}
    \label{fig:demand}
\end{figure}

The buildings connected to the network are four residential buildings in Cook County, Illinois, with data taken from NREL's ResStock \cite{horowitz2019resstock}. The heat capacities were estimated from the building's OpenStock models and are presented in \cref{tbl:build}. Their nominal heat demands are shown in \cref{fig:demand}. It was assumed that the maximum allowable temperature deviation was 2\degree C.\par
The chosen optimization objective for \cref{eq:i_cost} was
\begin{equation}
    f = \frac{w_{C}}{n_{U\elb{i}}}\left(\frac{SOE\elb{U\elb{i}}}{C\elb{U\elb{i}}\Delta T_b}\right)^2+w_Q hAs\left(T_P\elb{N\elb{i}}-T_{amb}\right)
\end{equation}
which is a weighted combination of the percent of used heat capacity in the subgraph and the heat lost by the network, where $w_C=5$ and $w_Q=3\times10^{-6}$.\par
The simulations were performed in discrete time with a 10-minute control time step, a 30-second discretization of temperature variables, a 1-hour optimization horizon, and a 10-minute implementation horizon. The simulation was performed over a 12-hour window starting on January 28th. All optimization was performed using CasADi \cite{Andersson2019} and IPOPT \cite{wachterImplementationInteriorpointFilter2006}.
\begin{table}
\caption{Parameters for network.}
\label{tbl:network}
\centering
\begin{tabular}{l c c c}
\toprule
Parameter & Symbol & Value & Units\\
\midrule
Supply temperature & $T_0$ & 80 & $C$\\
Ambient temperature & $T_{amb}$ & -16 -- -14 & $C$\\
Pipe diameter & $D$ & 0.15 -- 0.40 & $m$\\
Pipe length & $L$ & 20 -- 80 & $m$\\
Bypass length & $L$ & 3 & $m$\\
Heat transfer coefficient & $h$ & 1.5 & $W/m^2K$\\
Friction coefficient & $\lambda$ & 1 & -\\
Valve minimum & $\theta_{min}$ & 0.01 & -\\
Valve coefficient & $\mu$ & 2.6 & -\\
\bottomrule
\end{tabular}
\end{table}

\begin{table}
\caption{Parameters for buildings.}
\label{tbl:build}
\centering
\begin{tabular}{c c c c}
\toprule
$\mathcal{E}$ & Building ID & $C$ ($MJ/K$) & Initial \% $SOE$\\
\midrule
$e_4$ & R-3561 & 78 & 8\\
$e_{6}$ & R-80372 & 400 & -10\\
$e_{7}$ & R-3801 & 900 & 2\\
$e_{12}$ & R-80368 & 526 & -4\\
\bottomrule
\end{tabular}
\end{table}
\subsection{OLM Search Algorithm}\label{sec:bnb}
The graph partitioning problem is $\mathcal{NP}$-hard, meaning that in the worst-case, an exauhstive search is required to find the optimal solution. This paper uses a branch and bound (BnB) algorithm to solve \cref{eq:olm}. This algorithm relies on the hierarchical nature of graph partitioning to efficiently eliminate infeasible branches. The graph partitioning problem is considered a series of iterative bipartitions, where the newly-added group is bi-partitioned, allowing for a complete search of all partitions without repetition. In a system of $n_e$ elements, a minimum of $2^{n_e-1}$ initial bi-partitions must be explored, and the maximum number of possible partitions to be considered is given by Bell's number $B_{n_e}=\sum_{k=0}^{n_e-1}\binom{n_e-1}{k}B_k$.\par
In this BnB algorithm, first, all potential one-cut solutions are explored. For each solution ($\bar{p}$), two criteria are checked. If
\begin{equation}\label{eq:bound_new}
    olm(\bar{p})<olm(b)
\end{equation}
where $olm(b)$ is the cost of the best-found solution $b$, a new bound is established, and this branch is further explored. If this criterion is not met, then the condition
\begin{gather}
    \widetilde{olm}(\bar{p})<olm(b)\label{eq:bound_save}\\
    \widetilde{olm}=w_1c_{mPoA}+w_2c_{iter}+w_3\widetilde{c}_{sz}
\end{gather}
is considered where $\widetilde{c}_{sz}$ is the size of the largest partition that will not be further cut. If this criterion is met, then, while the bound is not updated, the branch is explored as further cutting could decrease $c_{sz}$ enough to improve on the current bound.\par
The branch exploration process is recursive. First, the elements assigned to the second group of the bi-partition are cut into two groups, and the $olm$ is calculated. Then \cref{eq:bound_new} and \cref{eq:bound_save} are evaluated. If \cref{eq:bound_save} is met, this process is repeated until \cref{eq:bound_save} is no longer met, and the branch is terminated, or the size of the newest group is one and the branch is fully explored. Note that these criteria assume that future partitioning can only increase the $mPoA$ and iterations until convergence. An outline of this procedure is presented in \cref{alg:bnb}.\par
\subsection{OLM Partitioning Results}
\begin{figure}
    \centering
    \begin{subfigure}{.5\linewidth}
        \centering
        \includegraphics[width=1\linewidth,trim={0 0 0 0},clip]{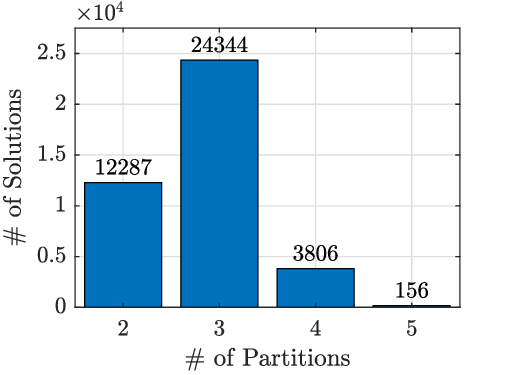}
        \caption{Solutions explored.}
        \label{fig:bnb_explored}
    \end{subfigure}%
    \begin{subfigure}{.5\linewidth}
        \centering
        \includegraphics[width=1\linewidth,trim={0 0 0 0},clip]{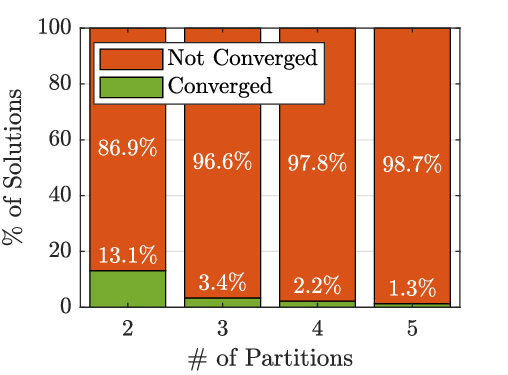}
        \caption{Percent converged.}
        \label{fig:bnb_converge}
    \end{subfigure}
    \caption{Solutions found and convergence rate by number of partitions.}
    \label{fig:bnb_stats}
\end{figure}

A single optimization step was used to calculate the OLM. The same initial guess, the solution to meet the nominal heat demand, was provided to the optimization solver for all $mPoA$ calculations for consistency. The chosen weights for \cref{eq:olm} were $w_1 = 1$, $w_2=0.04$, $w_3 = 0.06$. Each partition was given 20 iterations to converge. Partitions that did not converge in this horizon were eliminated. Additionally, divergent solutions were terminated early to save on computation time. 81,186 solutions were explored, split by number of partitions in \cref{fig:bnb_explored}. A system of 15 elements has $B_{15} \approx 1.4$ billion solutions, so the branch and bound algorithm was able to find the optimal solution considering only .006\% of the total search space. The algorithm only considered a maximum of five partitions (4 cuts) before all solution branches could be terminated. Only 6.2\% of the solutions converged, and the convergence percentage broken down by number of partitions is shown in \cref{fig:bnb_converge}.\par 

\begin{figure}\label{fig:part}
    \centering
    \begin{subfigure}{.5\linewidth}
        \centering
        \includegraphics[width=1\linewidth,trim={35 20 25 15},clip]{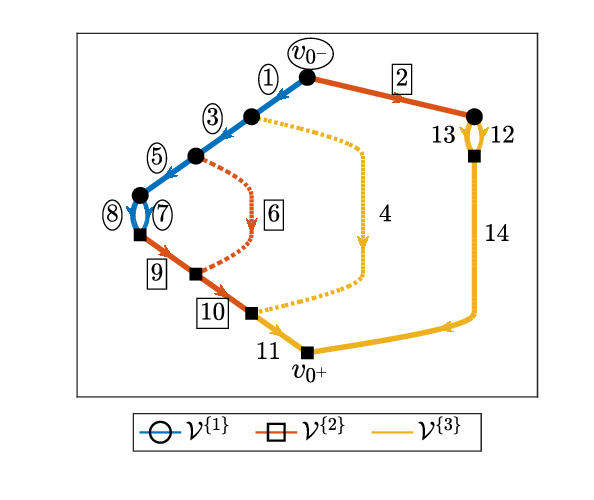}
        \caption{OLM-minimizing partition.}
        \label{fig:part_olm}
    \end{subfigure}%
    \begin{subfigure}{.5\linewidth}
        \centering
        \includegraphics[width=1\linewidth,trim={35 20 25 15},clip]{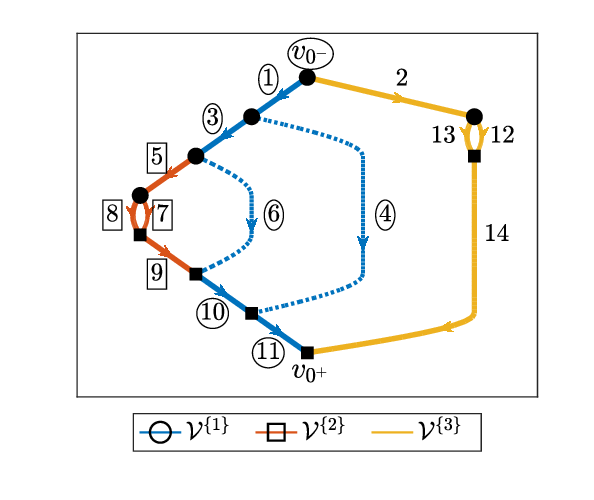}
        \caption{Baseline partition.}
        \label{fig:part_base}
    \end{subfigure}
    \caption{Partitioning results shown on flow graph.}
\end{figure}

The final OLM-minimizing partition is shown in \cref{fig:part_olm}. The $c_{mPoA} = 1.0030$ and this partition took two iterations to converge, the minimum possible, and the maximum size of a group in the partition is six elements. Note that partition two is disconnected, increasing its social awareness of the behaviors of partitions one and three. Partitions one and three contain the bypass elements, where waste flow is directed, a key factor in minimizing network losses. Additionally, all partitions contain at least one user element. The plant is controlled by partition one.

\subsection{Baseline Partitioning}
The results are compared to a baseline partitioning method that relies on a weighted graph representation of the system to indicate coupling between system elements. Here, a case-specific weight, $c_1$, is used, as this coefficient indicates heat capacity and therefore time delay in the pipes. The weighted graph is partitioned as follows. First, the original graph is converted to a line graph, where the pipes are represented as nodes, and the edges represent connections between the pipes, with the $c_1$ coefficient as weights. Then, the modularity maximizing partition is found \cite{jogwarCommunitybasedSynthesisDistributed2017}. The result of the baseline partition method for the case study is shown in \cref{fig:part_base}. In this graph, all partitions are connected and contain all sets of equivalent hot and cold edges. All partitions have at least one user, and partitions 2 and 3 contain bypass segments. There was no systematic way to determine the plant control via this partitioning method, so it was manually assigned to partition one.
\subsection{Simulation Results}
\begin{figure}
    \centering
    \begin{subfigure}{.5\linewidth}
        \centering
        \includegraphics[width=1\linewidth,trim={0 0 0 0},clip]{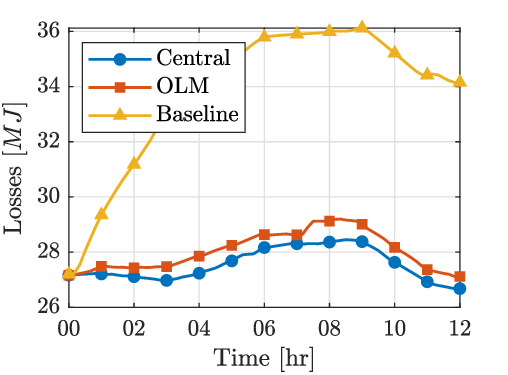}
        \caption{Energy losses.}
        \label{fig:losses}
    \end{subfigure}%
    \begin{subfigure}{.5\linewidth}
        \centering
        \includegraphics[width=1\linewidth,trim={0 0 0 0},clip]{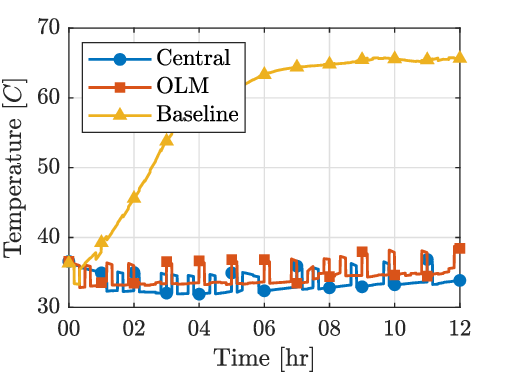}
        \caption{Return temperature.}
        \label{fig:Treturn}
    \end{subfigure}
    \caption{Losses and return temperature in all cases.}
    \label{fig:lossesT}
\end{figure}

\begin{figure}
    \centering
    \begin{subfigure}{.5\linewidth}
        \centering
        \includegraphics[width=\linewidth,trim={0 0 0 0},clip]{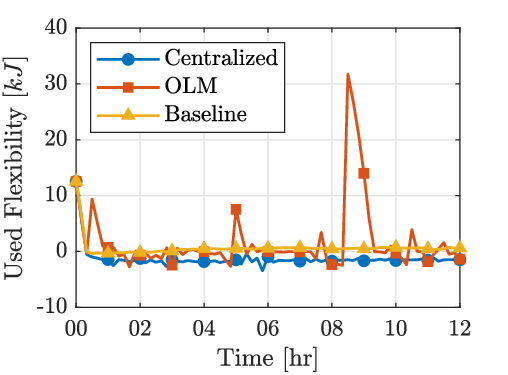}
        \caption{User $e_4$.}
        \label{fig:flex_4}
    \end{subfigure}%
    \begin{subfigure}{.5\linewidth}
        \centering
        \includegraphics[width=\linewidth,trim={0 0 0 0},clip]{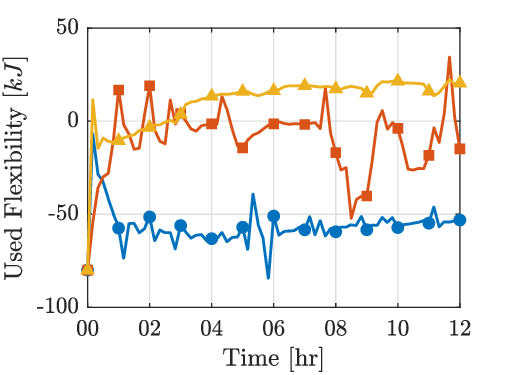}
        \caption{User $e_6$.}
        \label{fig:flex_6}
    \end{subfigure}
    \begin{subfigure}{.5\linewidth}
        \centering
        \includegraphics[width=\linewidth,trim={0 0 0 0},clip]{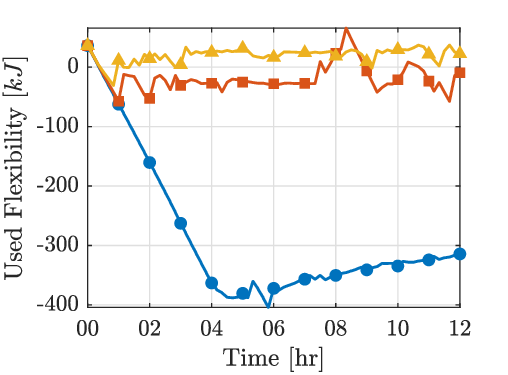}
        \caption{User $e_7$.}
        \label{fig:flex_7}
    \end{subfigure}%
    \begin{subfigure}{.5\linewidth}
        \centering
        \includegraphics[width=\linewidth,trim={0 0 0 0},clip]{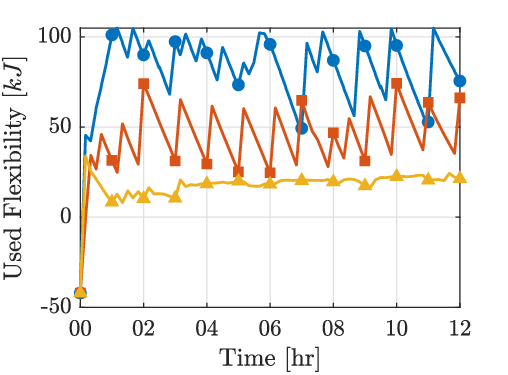}
        \caption{User $e_{12}$.}
        \label{fig:flex_12}
    \end{subfigure}
    \caption{Comparison of used flexibility by user.}
    \label{fig:flex}
\end{figure}

\begin{figure}
    \centering
    \includegraphics[width=.9\linewidth,trim={20 0 30 10},clip]{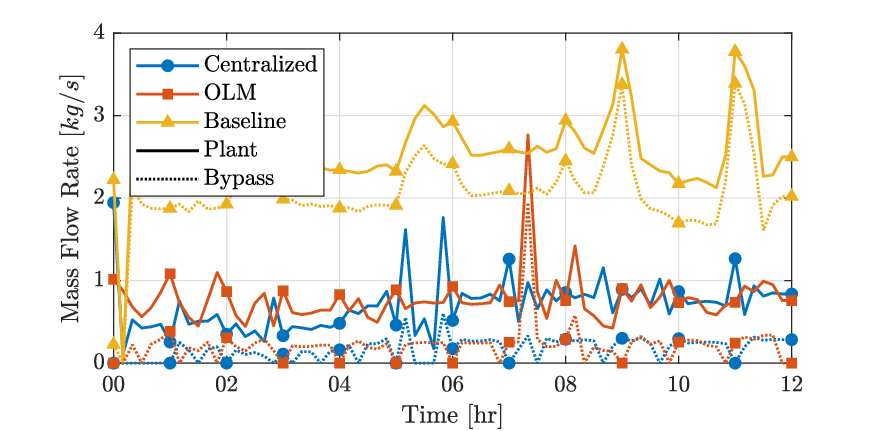}
    \caption{Plant and bypass flow in all cases.}
    \label{fig:plantflow}
\end{figure}

\begin{figure}
    \centering
    \begin{subfigure}{1\linewidth}
        \centering
        \includegraphics[width=1\linewidth,trim={20 0 35 5},clip]{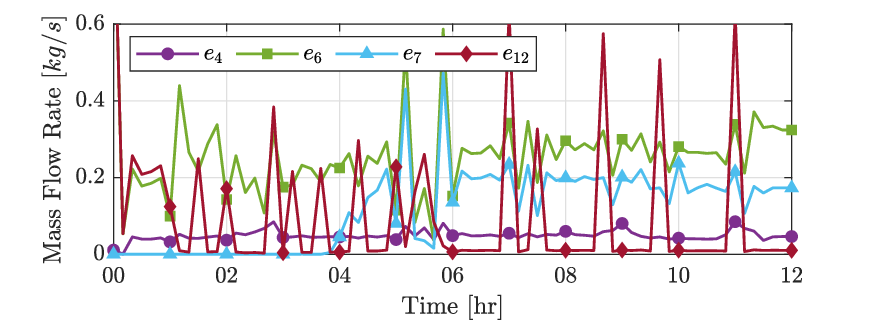}
        \caption{Centralized.}
        \label{fig:userflow_cen}
    \end{subfigure}
    \begin{subfigure}{1\linewidth}
        \centering
        \includegraphics[width=1\linewidth,trim={20 0 35 5},clip]{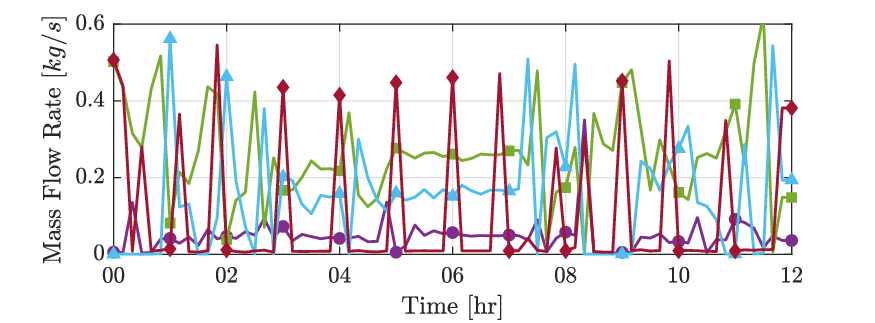}
        \caption{OLM.}
        \label{fig:userflow_olm}
    \end{subfigure}
    \begin{subfigure}{1\linewidth}
        \centering
        \includegraphics[width=1\linewidth,trim={20 0 35 5},clip]{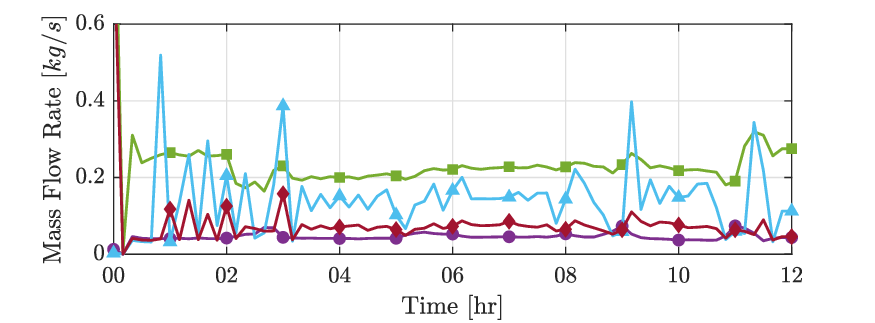}
        \caption{Baseline.}
        \label{fig:userflow_base}
    \end{subfigure}
    \caption{Comparison of flow delivered to the users.}
    \label{fig:userflow}
\end{figure}

After finding the OLM-minimizing and baseline partitions, a 12-hour control simulation was performed to compare the distributed performance to the centralized case. The results of these simulations are shown in \cref{fig:plantflow,fig:flex,fig:userflow,fig:lossesT}. The total losses, used SOE, and total costs for the three cases are summarized in \cref{tbl:cost}. Overall, the system partitioned using the novel performance-based method performed very similarly to the centralized solution, while the system partitioned using the baseline method had a large increase in losses.  \par 
\begin{table}
\caption{Cost components of the partitioning methods.}
\label{tbl:cost}
\centering
\begin{tabular}{c c c c}
\toprule
Method & Used Capacity (\%) & Losses (GJ) & Total Cost\\
\midrule
Centralized & 58.4 & 1.85 & $5.62\times 10^3$\\
OLM & 17.5 & 1.88 & $5.66\times 10^3$ \\
Baseline & 8.6 & 2.26 & $6.79\times 10^3$\\
\bottomrule
\end{tabular}
\end{table}

\Cref{fig:losses} shows the heat losses to the environment and return temperatures in the centralized and both distributed cases. This shows that while the return temperature in the OLM-partitioned case follows a similar stepped trend to the centralized one, correlated to periods of charging the SOE of the building, the return temperature in the baseline case is consistently high. Lower return temperature indicates a better use of the heat by the building, which decreases the temperature difference between the pipes and the environment and reduces overall losses, as shown in \cref{fig:Treturn}. The losses in the olm partitioned case increase only 1.7\% from the centralized case, versus 22.7\% in the baseline case.\par
\Cref{fig:flex} shows how the controller used the flexibility of the four buildings in the network. This figure shows that the centralized case, in general, is better able to take advantage of the flexibility in the buildings, with a higher proportion of the envelope being used. The baseline case has limited use of the flexibility, and the OLM case falls in between these two methods, consistent with their performance. Note that only a small percentage of the envelope was used in any of the simulations, never coming near the flexibility limits, as excessive deviation was penalized by the cost function.
Finally, \cref{fig:plantflow} and \cref{fig:userflow} show the mass flow rate from the plant, bypass segments, and user edges in the three cases. \Cref{fig:plantflow} shows that the difference between the bypass and plant flow is similar between the three cases, indicating a similar amount of flow delivered to the buildings. However, both the total and bypass mass flow are much lower in the centralized and OLM cases, and lower bypass flow correlates to the lower return temperature. The offset bypass flow in the baseline case indicates wasted flow. In all cases, the peaks in the bypass flow follow the trend of the plant flow. The lower bypass flow is further explained by \cref{fig:userflow}. In the centralized and olm cases, there are peaks and valleys in the users' flows, which allow for effective use of the flexibility of the buildings. These peaks and valleys are also staggered to allow for trade-offs between users receiving the flow. In the baseline case, the flow rate is more constant with only a few smaller peaks, and no points where the users received zero flow.
\section{Conclusion}\label{sec:conc}
This paper presents a novel partitioning method that uses performance degradation as a metric to determine the best partitioning for a distributed, communication-based, non-cooperative controller. By shifting from a system perspective to a control perspective in the design of the partition methods, this paper presents a novel partitioning metric that is more socially aware and better able to consider the control objectives and system constraints than previous methods. The metric used to find this partition is controller agnostic, meaning it is broadly applicable to many complex control applications, including DHNs. Overall, this novel method outperforms existing methods and enables the design of a distributed controller with near-optimal performance.\par
The effectiveness of this performance-based partitioning method is demonstrated on the distributed control of a four-user DHN, with proven stability and feasibility guarantees on the resulting controller. The automated nature of the control design makes it easily scalable to larger DHNs with more users. In simulation, the designed distributed controller shows little performance degradation compared to the centralized controller and significantly outperforms a baseline partitioning method.\par
Future work will address some of the limitations exposed by this method. First, future work will look at heuristic methods to find near-optimal approximations of the solution, as the $\mathcal{NP}$-hard graph partitioning problem will not scale efficiently to larger DHNs. Additionally, as operating conditions greatly impact system performance and the desired partitioning, future work will look at dynamic partitioning based on the SOE and environmental conditions. 
\bibliographystyle{IEEEtran}
\bibliography{sources}

\appendix
\input{notation}

\subsection{Stability Proof}\label{apx:stability}
The stability of any DHN partitioning will be proven here using passivity. A system is considered passive if it can not generate its own energy, and will only dissipate any initially stored energy \cite{arcakNetworksDissipativeSystems2016}. This proof is presented constructively. First, the passivity of an individual pipe is proven. Then, the passivity of the connection of pipes is proven. Finally, the stability of passive systems is ensured using zero-state detectability.

\subsubsection{Passivity of a Single Pipe}
In this framework, an individual pipe has four inputs, one output, and one state. The inputs are the negative temperature of the directly downstream pipes $-T_P\elb{i^+}$, the mass flow rate $\dot{m}_P\elb{i^+}$ of any directly downstream pipes, the power provided by the upstream pipe $\dot{Q}_{in}\elb{i}$, given by \cref{eq:Qin}, and power provided by the environment $\dot{Q}_{amb}\elb{i}$ given in \cref{eq:Qamb}.
The output of the system is the power provided to any directly downstream pipes, given by
\begin{equation}
    \dot{Q}_{out}\elb{i} = \dot{m}_P\elb{i^+}(T_P\elb{i}-T_P\elb{i^+})
\end{equation}
The system state is the current pipe temperature $T_P\elb{i}$ described by \cref{eq:Tpipe}.\par
The passivity of a single pipe is ensured via
\begin{equation}\label{eq:passpipe}
    \nabla_{T_P\elb{i}}V(T_P\elb{i})f(T_P\elb{i},u_i) <= s_i
\end{equation}
where $V(T_P\elb{i})$ is the storage function and $s_i$ is the supply rate. The input vector, $u_i$, is artificially constructed, containing all the original inputs as
\begin{equation}
    u_i = \begin{bmatrix}
        \dot{Q}_{in}\elb{i}&\dot{Q}_{amb}\elb{i} &\dot{m}_P\elb{i^+}c_pT_P\elb{i^+}& -T_P\elb{i^+}
    \end{bmatrix}^\top 
\end{equation}
The vector of outputs, $y_i$, is similarly constructed as
\begin{equation}
    y_i = \begin{bmatrix}
        T_P\elb{i} & T_P\elb{i} & T_P\elb{i} & \dot{Q}_{out}\elb{i}
    \end{bmatrix}^\top 
\end{equation} 
From the input and output vectors, the supply rate $s_i$ is calculated as
\begin{equation}
    s_i = u_i^\top y_i = T_P\elb{i}(\dot{Q}_{in}\elb{i}+\dot{Q}_{amb}\elb{i})+\dot{m}_P\elb{i^+}c_p{\left(T_P\elb{i^+}\right)}^2
\end{equation}
The storage function is defined as
\begin{equation}
    V(T_P\elb{i}) = \frac{1}{2}c_r\elb{i}{\left(T_P\elb{i}\right)}^2\geq0
\end{equation}
where $\smash{c_r\elb{i}=\left(\rho c_pV\right)\elb{i}}$.
combining these equations with \cref{eq:passpipe}, the passivity is ensured by
\begin{gather}
    \begin{multlined}
                c_r\elb{i}T_P\elb{i}\frac{1}{c_r\elb{i}} \left(\dot{Q}_{in}\elb{i} +\dot{Q}_{amb}\elb{i}\right)\leq\\
                T_P\elb{i}\left(\dot{Q}_{in}\elb{i} +\dot{Q}_{amb}\elb{i}\right) +\dot{m}_P\elb{i^+} c_p{\left(T_P\elb{i^+}\right)}^2
    \end{multlined}\\
    \dot{m}_P\elb{i^+}c_p{\left(T_P\elb{i^+}\right)}^2\geq 0\ \blacksquare
\end{gather}
where $f(T_P\elb{i},u_i)$ was replaced by the dynamics of the system state. As the mass flow rate is always positive due to how upstream and downstream are defined, this inequality holds, and the pipe is proven passive.
\subsubsection{Network Passivity}
\begin{figure}
    \centering
    \includegraphics[width=.7\linewidth]{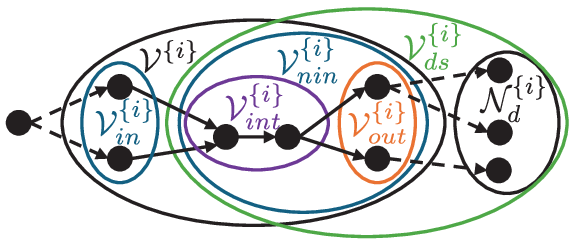}
    \caption{Visual representation of the described sets.}
    \label{fig:stabsets}
\end{figure}

Now considering a single partition $\nodelg{}\in \mathcal{G}$, five sets of nodes are defined
\begin{gather}
    \nodelg{in} = \left\{ v \in \nodelg{} \;\middle|\; \operatorname{indeg}(v) = 0 \right\}\\
    \nodelg{out} = \left\{ v \in \nodelg{} \;\middle|\; \operatorname{outdeg}(v) = 0 \right\}\\
    \nodelg{nin}=\nodelg{}\backslash\nodelg{in}\\
    \nodelg{int} = \nodelg{} \backslash \left(\nodelg{in} \cup \nodelg{out} \right)\\
    \nodelg{ds} = \left( \nodelg{} \backslash \nodelg{in} \right) \cup \mathcal{N}\elb{i}_d
\end{gather}
where $\nodelg{in}$ is the set of nodes feeding the partition, $\nodelg{nin}$ is its complement, $\nodelg{out}$ is the set of nodes exiting the partition, $\nodelg{int}$ are the nodes in the partition that are in neither $\nodelg{in}$ nor $\nodelg{out}$, and $\nodelg{ds}$ are the non-feeding nodes supplemented with the downstream neighbors outside the partition. These sets are illustrated in \cref{fig:stabsets}.
The inputs to this network are the negative downstream temperatures, the local and downstream mass flow rates, the power provided by upstream pipes, and the power provided by the environment. Respectively, these are given by:
\begin{equation*}
    -T_P\elb{\mathcal{N}\elb{i}_d}, \quad
    \dot{m}_P\elb{\nodelg{ds}}, \quad
    \dot{Q}_{in}\elb{\nodelg{in}}, \quad
    \dot{Q}_{amb}\elb{\nodelg{}}
\end{equation*}
The output is the power provided to the downstream pipes, $\dot{Q}_{out}\elb{\nodelg{out}}$, and the states are the local temperatures, $T_P\elb{\nodelg{}}$. Similar to the single pipe, the input matrix is constructed as
\begin{multline}
    u = \left[\begin{matrix}
        {\dot{Q}_{in}\elb{\nodelg{in}}}^\top&
        {\dot{Q}_{amb}\elb{\nodelg{}}}^\top\end{matrix}\right.\\
        \left.\begin{matrix}\left(c_p\dot{m}_P\elb{\nodelg{ds}}\odot\begin{bmatrix}
            T_P\elb{\nodelg{nin}}\\
            T_P\elb{\mathcal{N}\elb{i}_d}
        \end{bmatrix}\right)^\top&
        {-T_P\elb{\mathcal{N}\elb{i}_d}}^\top
    \end{matrix}\right]^\top
\end{multline}
and the output matrix is constructed as
\begin{multline}
    y =\left[ \begin{matrix}
    {T_P\elb{\nodelg{in}}}^\top&
    {T_P\elb{\nodelg{}}}^\top\end{matrix}\right.\\
    \left.\begin{matrix}
    \begin{bmatrix}
        {T_P\elb{\nodelg{int}}}^\top &
        {T_P\elb{\nodelg{out}}}^\top
    \end{bmatrix}&
    {\dot{Q}_{out}\elb{\nodelg{out}}}^\top
\end{matrix}\right]^\top
\end{multline}
The state dynamics equation is written as
\begin{equation}
    \frac{d}{dt}\begin{bmatrix}
        T_P\elb{\nodelg{in}}\\
        T_P\elb{\nodelg{nin}}
    \end{bmatrix}=
    \begin{bmatrix}
        \frac{1}{c_r\elb{\nodelg{in}}}\odot\left(\dot{Q}_{in}\elb{\nodelg{in}}+\dot{Q}_{amb}\elb{\nodelg{in}}\right)\\
        \begin{multlined}
             \frac{1}{c_r\elb{\nodelg{nin}}}\odot\left(c_p
        \dot{m}_P\elb{\nodelg{nin}}\odot\right.\\
        \left.\left(T_P\elb{\nodelg{int}}-T_P\elb{\nodelg{nin}}\right) + \dot{Q}_{amb}\elb{\nodelg{nin}}\right)
        \end{multlined}
    \end{bmatrix}
\end{equation}
and the storage function as
\begin{equation}
    \frac{1}{2}{c_r\elb{\nodelg{}}}^\top T_P\elb{\nodelg{}}
\end{equation}
Then passivity is proven by showing $\nabla_{T_p\elb{\nodelg{}}}V\left(T_p\elb{\nodelg{}}\right) F\left(T_{\nodelg{}},u\right) \leq u^\top y$ in \cref{eq:pass_proof}
\begin{figure*}
\begin{equation}\label{eq:pass_proof}
\begin{gathered}
        \begin{multlined}
            {T_P\elb{\nodelg{in}}}^\top  \left(\dot{Q}_{in}\elb{\nodelg{in}} + \dot{Q}_{amb}\elb{\nodelg{in}} \right) +
            {T_P\elb{\nodelg{nin}}}^\top  \left(c_p\dot{m}_P\elb{\nodelg{nin}} \odot \left(T_P\elb{\nodelg{int}} - T_{P}\elb{\nodelg{nin}}\right) + \dot{Q}_{amb}\elb{\nodelg{nin}} \right) \leq\\
           {\dot{Q}_{in}\elb{\nodelg{in}}}^\top T_P\elb{\nodelg{in}} + {\dot{Q}_{amb}\elb{\nodelg{}}}^\top T_P\elb{\nodelg{}} +
           \left(\ c_p\dot{m}_P\elb{\nodelg{ds}}\odot \begin{bmatrix}
               T_P\elb{\nodelg{nin}}\\T_P\elb{\mathcal{N}\elb{i}_d}
           \end{bmatrix} \right)^\top \begin{bmatrix}
               T_P\elb{\nodelg{int}}\\T_P\elb{\nodelg{out}}
           \end{bmatrix}-T_P\elb{\mathcal{N}\elb{i}_d}\dot{Q}_{out}\elb{\nodelg{out}}
        \end{multlined}\\
        -c_p \dot{m}_P\elb{\nodelg{nin}} \odot {T_P\elb{\nodelg{nin}}}^2 \leq c_p \dot{m}_P\elb{\mathcal{N}\elb{i}_d} \odot {T_P\elb{\mathcal{N}\elb{i}_d}}^2\ \blacksquare
\end{gathered}
\end{equation}
\end{figure*}
where, by definition, the mass flow is always positive, the inequality holds, and the interconnection of the passive pipe elements is also passive.
\subsubsection{Stability via Passivity}
Passive systems are stable if and only if they are zero-state detectable (ZSD) \cite{sepulchreConstructiveNonlinearControl1997}. ZSD indicates that when the system's inputs are zero, the system's states are also ensured to be zero after some time. For the system $H$ with zero inputs ($\dot{x} = f(x,0),y=h(x,0)$) and its largest positively invariant set $Z \subset \mathbb{R}^n$ contained in $\left\{x\in\mathbb{R}^n|y=h(x,0)=0\right\}$, $H$ is ZSD if $x=0$ is asymptotically stable conditionally to $Z$. If $Z=0$, $H$ is said to be zero-state observable (ZSO).\par
For the single pipe, when no heat is injected, assuming the equilibrium point is shifted to zero, 
\begin{gather}
    \frac{d}{dt}T_P\elb{i} = 0\\
    y_i^{red} = \begin{bmatrix}
        T_P\elb{i}& \dot{Q}_{out}\elb{i}
    \end{bmatrix}^\top = \begin{bmatrix}
        T_P\elb{i}&
        \dot{m}_P\elb{i^+}c_pT_P\elb{i}
    \end{bmatrix}^\top
\end{gather}
where $y_i^{red}$ is the output vector, reduced to eliminate redundancies. Then, $y_i^{red}=0$ only if $T_P\elb{i} = 0$ indicating the pipe is ZSO and therefore ZSD.\par
For the network of pipes, assuming no heat is injected
\begin{gather}
        \frac{d}{dt}\begin{bmatrix}
        T_P\elb{\nodelg{in}}\\
        T_P\elb{\nodelg{nin}}
    \end{bmatrix}=
    \begin{bmatrix}
        0\\
        \begin{multlined}
            \frac{1}{c_r\elb{\nodelg{nin}}}\odot\left(c_p
        \dot{m}_P\elb{\nodelg{nin}}\odot\right.\\
        \left.\left(T_P\elb{\nodelg{int}}-T_P\elb{\nodelg{nin}}\right) + 0\right) \end{multlined}
    \end{bmatrix}\\
    y^{red} = \begin{bmatrix}
        T_P\elb{\nodelg{}}\\ \dot{Q}_{out}\elb{\nodelg{out}}
    \end{bmatrix} = \begin{bmatrix}
        T_P\elb{\nodelg{}}\\
        \dot{m}_P\elb{i^+}c_p \odot \left( T_P\elb{\nodelg{out}}-0 \right)
    \end{bmatrix}
\end{gather}
where $y^{red}$ is also reduced to eliminate redundancies. Then, $y^{red}=0$ only if $T_P\elb{\nodelg{}} = 0$, indicating the interconnected system is ZSO and therefore ZSD. Finally, as both the individual elements of the system and their interconnections are both passive and ZSD, any partitioning of the original system results in stable subsystems, ensuring stability will not be a factor in the system partitioning. 

\subsection{Branch and Bound Algorithm}
\Cref{alg:bnb} is pseudocode for the algorithm described in \cref{sec:bnb}. Here, $p_n$ indicates all potential bi-partitions of $n$ elements with the newest group arbitrarily assigned, $tbe$ is a running list of partitions to be branched, $\operatorname{szn}\left(\bar{p}\right)$ indicates the size of the newest group in $\bar{p}$, and $\bar{p}\cup \widetilde{p}$ indicates replacing the elements in the newest group of $\bar{p}$ with $\widetilde{p}$. Finally, \cref{algl:check} ensures that the plant remains in a partitioning with one of its directly connected edges. Finding the optimality loss metric for each solution in $tbe_i$ is parallelizable to enable faster solving. 
\begin{algorithm}
    \caption{Branch and bound for graph partitioning.}
    \label{alg:bnb}
    \begin{algorithmic}[1]
    \State{$olm(b)\gets \infty$, $tbe_1\gets \mathbf{1}^{n_{e}}$}
    \For{$n=1\dots n_{e}-1$}
        \State{Initialize $s$}
        \For{$i\ =1\dots \operatorname{card}\left(tbe_{n}\right)$}
            \State{$\bar{s}\gets$ \Call{Branch}{$tbe_{n}\elb{i}$}}
            \State{Append $\bar{s}$ to $s$}
        \EndFor
        \If{$\min_{i}\left(olm\left(s\elb{i}\right)\right)<olm\left(b\right)$}
            \State{$b\gets s\elb{i}$ s.t. $i =\operatorname{argmin}_i \left( olm\left( s\elb{i} \right) \right)$}
        \EndIf
        \State{$tbe_{n+1}\gets s\elb{i} \text{ s.t. } \widetilde{olm}\left(s\elb{i}\right)\leq olm(b)$}
    \EndFor
    \State{\textbf{return} $b$}
    \Procedure{Branch}{$\bar{p}$}
    \State{$n_n \gets \operatorname{szn}\left(\bar{p}\right)$}
    \If{$n_n>1$}
        \State{Initialize $s$}
        \For{$i = 1 \dots 2^{n_n-1}$}
            \If{$\bar{p}\cup p_{n_n}\elb{i}$ is valid}\label{algl:check}
                \State{Append $\bar{p}\cup p_{n_n}\elb{i}$ to $s$}
            \EndIf
        \EndFor
    \Else
        \State{End branch, $s\gets \emptyset$}
    \EndIf
        
    \State{\textbf{return} $s$}
    \EndProcedure
    
    \end{algorithmic}
\end{algorithm}
\end{document}

%% file: notation.tex
\subsection{Notation}\label{sec:notation}
\subsubsection*{Variables}
\begin{itemize}
    \item[$\mathcal{G}$] Graph
    \item[$\mathcal{V}$] Node set
    \item[$\mathcal{E}$] Edge set
    \item[$\mathcal{N}$] Neighbor
    
    \item[$F$] Feeding pipes
    \item[$R$] Return pipes
    \item[$By$] Bypass pipes
    \item[$U$] User elements
    \item[$N$] Non-user pipes
    \item[$\Gamma$] Adjacency matrix of $\mathcal{G}$
    \item[$\Lambda$] Incidence matrix of $\mathcal{G}_f$

    \item[$\dot{m}$] Mass flow rate
    \item[$\theta$] Valve position
    \item[$T$] Temperature
    \item[$\dot{Q}$] Heat transfer
    \item[$\rho$] Density
    \item[$c_p$] Specific heat capacity
    \item[$V$] Volume
    \item[$hAs$] Heat transfer coefficent
    \item[$P$] Pressure
    \item[$\zeta$] Pressure loss coefficient
    \item[$\mu$] Valve scaling coefficient
    \item[$SOE$] State of energy
    \item[$C$] Heat storage coefficients
    
\end{itemize}

\subsubsection*{Subscripts}
\begin{itemize}
    \item[$f$] Flow
    \item[$0$] Plant
    \item[$amb$] Ambient
    \item[$P$] Pipe
    \item[$S$] Split
\end{itemize}